\documentclass[journal]{IEEEtranTIE}

\usepackage{graphicx}
\usepackage{cite}
\usepackage{amsmath}
\usepackage{amssymb}
\usepackage{url}
\usepackage{colortbl}
\usepackage{soul}
\usepackage{multirow}
\usepackage{pifont}
\usepackage{color}
\usepackage{alltt}
\usepackage{enumerate}
\usepackage{epstopdf}
\usepackage{pbox}
\usepackage[caption=false,font=footnotesize]{subfig}
\usepackage{placeins}
\usepackage{algorithm}
\usepackage{algpseudocode}
\usepackage{setspace}

\DeclareSymbolFont{matha}{OML}{txmi}{m}{it}
\DeclareMathSymbol{\varv}{\mathord}{matha}{118}

\begin{document}
\title{\huge{Real-Time Electromagnetic Estimation\\for Reluctance Actuators}}

\author{
	\vskip 1em
	Edgar Ramirez-Laboreo, Eduardo Moya-Lasheras, and Carlos~Sagues

	\thanks{

		Manuscript received September 28, 2017; revised February 2, 2018; accepted May 4, 2018.
		This work was supported in part by the Ministerio de Econom\'ia y Competitividad, Gobierno de Espa\~na - European Union, under project RTC-2014-1847-6, in part by the Ministerio de Educaci\'on, Cultura y Deporte, Gobierno de Espa\~na, under grant FPU14/04171, and in part by project DGA-T45\_17R/FSE.
		
		The authors are with the Departamento de Informatica e Ingenieria de Sistemas (DIIS) and the Instituto de Investigacion en Ingenieria de Aragon (I3A), Universidad de Zaragoza, Zaragoza 50018, Spain (e-mail: ramirlab@unizar.es, emoya@unizar.es, csagues@unizar.es).
	}
    \thanks{\textcolor{red}{This is the accepted version of the manuscript: E. Ramirez-Laboreo, E. Moya-Lasheras and C. Sagues, ``Real-Time Electromagnetic Estimation for Reluctance Actuators," in IEEE Transactions on Industrial Electronics, vol. 66, no. 3, pp. 1952-1961, March 2019, doi: 10.1109/TIE.2018.2838077 
    \textbf{Please cite the publisher's version}. For the publisher's version and full citation details see:\\\protect\url{https://doi.org/10.1109/TIE.2018.2838077}. 
}}
 \thanks{© 2018 IEEE.  Personal use of this material is permitted.  Permission from IEEE must be obtained for all other uses, in any current or future media, including reprinting/republishing this material for advertising or promotional purposes, creating new collective works, for resale or redistribution to servers or lists, or reuse of any copyrighted component of this work in other works.}
}

\maketitle
	
\begin{abstract}
Several modeling, estimation, and control strategies have been recently presented for simple reluctance devices like solenoid valves and electromagnetic switches. In this paper, we present a new algorithm to online estimate the flux linkage and the electrical time-variant parameters of these devices, namely the resistance and the inductance, only by making use of discrete-time measurements of voltage and current. The algorithm, which is robust against measurement noise, is able to deal with temperature variations of the device and provides accurate estimations during the motion of the armature. Additionally, an integral {estimator} that uses the start of each operation of the actuator as reset condition has been also implemented for comparative purposes. The performances of both estimation methods are studied and compared by means of simulations and experimental tests, and the benefits of our proposal are emphasized. Possible uses of the estimates and further modeling developments are also described and discussed.
\end{abstract}


\markboth{}
{}

\section{Introduction}

\IEEEPARstart{S}{imple} non-latching reluctance devices based, e.g., on plunger-type, pivoted-armature, or E-core actuators, are being increasingly used in several domains mainly because of their low cost. Thus, while the automotive industry has recently found novel uses for solenoid valves~\cite{Zhao2016linear}, electromagnetic switches can be widely found in many present-day applications, e.g., wireless power transfer systems~\cite{beh2013automated}, battery chargers for electric vehicles~\cite{haghbin2013grid}, or photovoltaic modules~\cite{sharma2014novel}.

As stated in some previous works \cite{Gaeta2008modeling,ramirez2016new}, the dynamics of these devices is governed by an electromagnetic force that increases greatly when the air gap is near zero. This nonlinear behavior, together with physical bounds that limit the motion, causes switches and valves to be subject to strong shocks and wear that often result in early failures. In order to overcome these problems and improve the performance of the devices, several control strategies have been presented. See, e.g., the current-limiting method in~\cite{dos2008electronic}, the energy-based approach in~\cite{yang2013multiobjective}, or some iterative approaches~\cite{Tsai2012cycle,ramirez2017new}. The major problem when controlling the motion is that the position of the mover with respect to the stator cannot be measured \mbox{-- at} least not with affordable \mbox{sensors --} and therefore feedback control can only be applied via estimation techniques. Some works can be found concerning the position estimation, e.g., a nonlinear sliding-mode observer is included in~\cite{Zhao2016linear} and a fuzzy controller is described in~\cite{espinosa2008sensorless}. The underlying idea of these proposals is that the inductance of the device depends on the position of the armature. Hence, if an accurate model relating these two variables were available and an inductance estimator implemented, the motion of the device may also be estimated.

In this paper, we propose a novel algorithm for stochastic electromagnetic estimation in reluctance actuators (SEMERA) which is able to estimate online the inductance, the resistance and the flux linkage of these devices, as well as additional variables, only by using discrete-time measurements of voltage and current. Apart from the inductance, which may be used to estimate the position of the mover, the resistance estimation can be used, e.g., as a temperature sensor of the device. On the other hand, the flux linkage may allow for estimating the magnetic force that drives the motion \cite{ramirez2016new} or for detecting magnetic hysteresis and saturation~\cite{vaughan1996modeling}. The observer is based on the celebrated Kalman filter theory~\cite{kalman1960new} and, in contrast to some recent approaches~\cite{mackenzie2016real}, it relies only on a simple model {of a variable inductor} that is not dependent on the position of the armature. {Besides, it includes a confidence interval (CI) evaluation method that detects the instants of low signal-to-noise ratio (SNR) and an expert rule that assigns values to the estimated variables during these periods.} On the other hand, an efficient integral {estimator}, whose reset condition is based on the cyclic operation of these devices, has been also developed for comparative purposes. The algorithms have been validated by simulation and then applied to two actual devices by means of a microcontroller-based prototype. Both simulation and experimental results are presented and analyzed.

The main contributions of the work are: (I)~a robust observer that, {without requiring any model of the device,} estimates the resistance, the inductance, and the flux linkage of reluctance actuators during the motion, even in the presence of measurement noise or temperature changes, (II)~an extension of the stochastic filtering theory for online estimation of variables and time-variant parameters, (III)~an observability analysis of the filter that provides insight into the system excitation and justifies the selection of state variables, and (IV)~a comparison between the algorithm and an \mbox{ad hoc} integral {estimator} based on the cyclic operation of switches and solenoid valves.

\section{Algorithm formulation}

As already stated, the SEMERA algorithm proposed in this paper is based on the Kalman filter theory. In this section we present the model used by the filter and provide an analysis of the resulting equations in terms of observability. We will use the notations $\hat{x}_{k/k-1}$ and $\hat{x}_{k/k}$ to refer, respectively, to the \textit{a priori} and \textit{a posteriori} estimates of the state $x$ at step $k$.

\subsection{Observation Model and Process Model}

The observation model of the filter is based on the dynamic equation of an inductor with internal resistance,
\begin{equation}
\varv(t) = r(t)i(t) + \frac{\mathrm{d}\lambda(t)}{\mathrm{d}t},
\label{eq:coil_cont}
\end{equation}
where $\varv(t)$ is the voltage across its terminals, $r(t)$ is the internal electrical resistance, $i(t)$ is the electrical current, and $\lambda(t)$ is the flux linkage. Although $r$ may be considered constant, it has been assumed time-dependent in order to account for temperature changes during the operation \cite{roemer2015optimum}. {This continuous-time equation is discretized by backward differentiation,}
\begin{equation}
\varv_k = r_k i_k + \frac{\lambda_k - \lambda_{k-1}}{\Delta},
\label{eq:coil_discrete_1}
\end{equation}
where $\Delta$ is the sampling period and the subscripts are used to indicate the time step. {First-order forward and central difference formulas may be used as alternatives for discretizing {\eqref{eq:coil_cont}}, but they result in a \mbox{one-step} delay in the estimation of $r$. On the other hand, higher order backward-differentiation expressions could also be utilized, but at the expense of increasing the order of the filter and the complexity of the model.}

For non-latching devices, i.e., devices without permanent magnets, $\lambda$ can be expressed as the product of the apparent inductance $l$ and the electrical current, $\lambda=li$. Hence, the previous equation is transformed into
\begin{equation}
\varv_k = r_k i_k + \frac{l_k i_k - l_{k-1} i_{k-1}}{\Delta},
\label{eq:coil_discrete_2}
\end{equation}
where the inductance $l$ is considered {a time-dependent variable} because, in any reluctance-based device, it changes with the motion of the mechanism. {A different discrete version of {\eqref{eq:coil_cont}} may be obtained if the derivative of $\lambda$ is first expanded,}
\begin{equation}
\varv(t) = r(t)i(t) +  \frac{\mathrm{d}l(t)}{\mathrm{d}t}i(t) + l(t)\frac{\mathrm{d}i(t)}{\mathrm{d}t},
\end{equation}
{and then the derivatives of $l$ and $i$ are replaced by their backward discrete approximations.} {Although in this paper we will use {\eqref{eq:coil_cont}} to derive the equations of the filter, it can be shown that both approximations have discretization errors ${O}(\Delta)$ and provide similar results.}

Experimental measurements of voltage and current are required by the filter at each time step. Since measurement processes always add noise to the actual variables, let us define the voltage observation, $u$, and the current observation, $\iota$, as
\begin{align}
u_k &= \varv_k + {v_\varv}_k, \label{eq:eps_k}\\
\iota_k &= i_k + {v_i}_k, \label{eq:y_k}
\end{align}
where ${v_\varv}$ and ${v_i}$ are additive noises that affect, respectively, the voltage measurement and the current measurement. Hence, combining \mbox{\eqref{eq:coil_discrete_2}-\eqref{eq:y_k}} and reorganizing terms, we obtain
\begin{align}
u_k = & \ \iota_k r_k + \dfrac{\iota_k l_k -\iota_{k-1} l_{k-1}}{\Delta} + \nonumber\\
& +{v_\varv}_k - {v_i}_k\left(r_k + {l_k}/{\Delta}\right) +  {v_i}_{k-1}{l_{k-1}}/{\Delta}.
\label{eq:epsilon_k}
\end{align}

It is easy to see now that \eqref{eq:epsilon_k} can be used as the observation equation of the filter,
\begin{equation}
z_k = H_k x_k + v_k,
\label{eq:kalman_output}
\end{equation}
where $z_k$ is the observed output, $H_k$ is the observation matrix, $x_k$ is the filter state vector, and $v_k$ is the observation noise at time step $k$, simply by selecting these variables as
\begin{align}
z_k &= u_k, \\
x_k &= \left[\begin{array}{ccc}r_k & l_k & l_{k-1}\end{array}\right]^\mathsf{T}, \label{eq:state}\\[1pt]
H_k &= \left[\begin{array}{ccc} \iota_k & \iota_k/\Delta & -\iota_{k-1}/\Delta \end{array}\right], \label{eq:H_k} \\[1pt]
v_k &= {v_\varv}_k - {v_i}_k\left(r_k + {l_k}/{\Delta}\right) +  {v_i}_{k-1}{l_{k-1}}/{\Delta}. \label{eq:v_k}
\end{align}
{This structure may resemble the equations used for real-time identification of autoregressive models}~\cite{anderson2005optimal}, {but note that the elements of $x_k$ are not independent parameters because $l_k$ and $l_{k-1}$ are time-connected. Note also that the observation noise $v_k$ depends on the state and may be rewritten as}
\begin{equation}
v_k = {v_\varv}_k - {V_i}_k x_k,
\label{eq:v_k2}
\end{equation}
\begin{equation}
{V_i}_k = \left[\begin{array}{ccc} {v_i}_k & {v_i}_k/\Delta & -{v_i}_{k-1}/\Delta \end{array}\right].
\end{equation}
{Then,} assuming that $\{{v_\varv}_k\}$ and $\{{v_i}_k\}$ are independent random processes with zero mean and known variances, $\mathrm{var}({v_\varv}_k) = \sigma^2_\varv$ and $\mathrm{var}({v_i}_k) = \sigma^2_i$, it can be shown that $\{v_k\}$ is also a zero-mean process with variance given by
\begin{equation}
R_k = \mathrm{var}\left(v_k\right) = \sigma^2_\varv + x_k^\mathsf{T} \left[\begin{array}{ccc}
\sigma^2_i & \dfrac{\sigma^2_i}{\Delta} & 0 \\[0.6em]
\dfrac{\sigma^2_i}{\Delta} & \dfrac{\sigma^2_i}{\Delta^2} & 0 \\[0.6em]
0 & 0 & \dfrac{\sigma^2_i}{\Delta^2}
\end{array}\right] x_k,
\end{equation}

On the other hand, the process model used by the filter is
\begin{equation}
x_{k+1} = F x_k + G w_k,
\label{eq:state_eq}
\end{equation}
where $F$ and $G$ are the discrete-time state and input matrices with proper dimensions, and $w_k$ is the input, or process, noise. This structure leads to the prediction model
\begin{equation}
\hat{x}_{k+1/k} = F \hat{x}_{k/k}.
\end{equation}

Given the dynamic behavior of the system, {we propose to approximate $r$ as a constant parameter and $l$ as a variable having a linear evolution in time}.
This leads to the predictions
\begin{align}
{\hat{x}_{k+1/k}}^{(1)} &= \hat{r}_{k+1/k} \approx \hat{r}_{k/k} = {\hat{x}_{k/k}}^{(1)},\\
{\hat{x}_{k+1/k}}^{(2)} &= \hat{l}_{k+1/k} \approx \hat{l}_{k/k} + \left(\hat{l}_{k/k}-\hat{l}_{k-1/k}\right) = \nonumber \\
&= 2{\hat{x}_{k/k}}^{(2)} - {\hat{x}_{k/k}}^{(3)}, \\
{\hat{x}_{k+1/k}}^{(3)} &= \hat{l}_{k/k} = {\hat{x}_{k/k}}^{(2)},
\end{align}
and, consequently, to a state transition matrix as follows
\begin{equation}
F = \left[\begin{array}{ccc}1 & 0 & 0 \\ 0 & 2 & -1 \\ 0 & 1 & 0\end{array}\right].
\label{eq:F}
\end{equation}

{Note that this model differs from those usually used in adaptive Kalman filtering} \cite{chui2017kalman}. {Apart from not assuming a constant inductance, the main difference is that our process model does not include the dynamics of the actual system; the only equation linking the filter to the actuator is the observation equation. In this way, the algorithm can be applied to any variable reluctance device independently of its particular design.}

Substituting \eqref{eq:F} in \eqref{eq:state_eq}, solving for $G w_k$, and approximating according to the Taylor series, we obtain the expression for the input term of the process model,
\begin{equation}
G w_k = \left[\begin{array}{c} r_{k+1}-r_k \\ l_{k+1}-2l_k+l_{k-1} \\ 0 \end{array}\right] \approx \left[\begin{array}{c} \dot{r}_k\,\Delta \\ \ddot{l}_k\,\Delta^2 \\ 0 \end{array}\right],
\end{equation}
where $\dot{r}$ and $\ddot{l}$ are, respectively, the first derivative of $r$ and the second derivative of $l$ with respect to time. Hence, in order to distinguish between constants and variables, $w_k$ and $G$ are selected as
\begin{align}
w_k &= \left[\begin{array}{c} \dot{r}_k \\ \ddot{l}_k \end{array}\right], &
G &= \left[\begin{array}{cc} \Delta & 0 \\ 0 & \Delta^2 \\ 0 & 0 \end{array}\right].
\end{align}
Then, assuming that $\{\dot{r}_k\}$ and $\{\ddot{l}_k\}$ are independent, zero-mean random processes with known variances, $\mathrm{var}\left(\dot{r}_k\right) = \sigma^2_{\dot{r}}$ and $\mathrm{var}(\ddot{l}_k) = \sigma^2_{\ddot{l}}$, the covariance matrix of the process noise, $Q$, is given by
\begin{equation}
Q = \left[\begin{array}{cc} \sigma^2_{\dot{r}} & 0 \\ 0 & \sigma^2_{\ddot{l}} \end{array}\right].
\end{equation}

Finally, let us assume that the initial values of resistance and inductance, $r_0$ and $l_0$, are also random processes with known expected values, $\mathrm{E}\left(r_0\right)=\bar{r}_0$ and $\mathrm{E}\left(l_0\right)=\bar{l}_0$, and known variances, $\mathrm{var}(r_0) = \sigma^2_{r_0}$ and $\mathrm{var}(l_0) = \sigma^2_{l_0}$. Hence, considering that $l_{-1}=l_{0}$, the expected value of the initial state, $\bar{x}_0$, and the initial covariance matrix, $P_0$, are given by
\begin{equation}
\bar{x}_0 = \mathrm{E}\left(x_0\right) = \left[\begin{array}{ccc}\bar{r}_0 & \bar{l}_0 & \bar{l}_0\end{array}\right]^\mathsf{T},
\end{equation}
\begin{equation}
P_0 = \mathrm{var}\left(x_0\right) = \left[\begin{array}{ccc}
\sigma^2_{r_0} & 0 & 0 \\[1pt]
0 & \sigma^2_{l_0} & \sigma^2_{l_0} \\[1pt]
0 & \sigma^2_{l_0} & \sigma^2_{l_0}
\end{array}\right].
\end{equation}

\subsection{Observability and Convergence}
\label{subsec:observability}

The observability of the proposed model has to be analyzed to confirm the feasibility of the estimator. In this regard it should be noted that, since observability is a structural property,
in this case it cannot be analyzed through the observation equation of the filter, \eqref{eq:kalman_output}, because $H_k$ depends on the measurement noise.
Instead, the \textit{structural} output equation,
\begin{equation}
y_k = C_k x_k,
\label{eq:output_struct}
\end{equation}
where $y_k$ is the true output (not to be confused with the observation, $z_k$) and $C_k$ is the output matrix at step $k$, has to be considered.
Given that the model output is the voltage through the coil, $y_k = \varv_k$, and that the state vector has been already selected in \eqref{eq:state}, the output matrix is obtained from~\eqref{eq:coil_discrete_2}~as
\begin{equation}
C_k = \left[\begin{array}{ccc} i_k & i_k/\Delta & -i_{k-1}/\Delta \end{array}\right].
\label{eq:C_k}
\end{equation}
{Note that, according to {\eqref{eq:y_k}} and {\eqref{eq:H_k}},} 
\begin{equation}
H_k = C_k + {V_i}_k.
\label{eq:H_k2}
\end{equation}

Now, since the model \eqref{eq:state_eq}, \eqref{eq:output_struct} is linear, the observability can be analyzed by means of the observability matrix. Strictly speaking, the presented time-variant model is \textit{observable} on the interval $t\in\left[\,k\Delta,\ (k+n)\Delta\,\right]$ if and only if the matrix
\begin{equation}
\mathcal{O}_{\left[k,\,k+n\right]} = \left[\begin{array}{c}
C_k \\ C_{k+1}F \\ ... \\ C_{k+n}F^{n}
\end{array}\right]
\end{equation}
is full rank. Given the arbitrary size of the previous matrix, let us analyze the observability on the interval $t\in\left[\,k\Delta,\ (k+2)\Delta\,\right]$, which, given the size of the state vector, is the shortest possible interval of observability. In this case, the observability matrix is given by 
\begin{equation}
\mathcal{O}_{\left[k,\,k+2\right]} \!=\! \left[\begin{array}{c@{\hspace{0.75em}}c@{\hspace{0.75em}}c}
i_k & \dfrac{i_k}{\Delta} & -\dfrac{i_{k-1}}{\Delta} \\[1em]
i_{k+1} & \dfrac{2i_{k+1}-i_k}{\Delta} & -\dfrac{i_{k+1}}{\Delta} \\[1em]
i_{k+2} & \dfrac{3i_{k+2}-2i_{k+1}}{\Delta} & \dfrac{i_{k+1}-2i_{k+2}}{\Delta} 
\end{array}\right],
\end{equation}
and the model is observable provided that the determinant,
\begin{align}
\mathrm{det}&\left(\mathcal{O}_{\left[k,\,k+2\right]}\right) = \big( 2i_{k-1}{i_{k+1}}^2+2{i_{k}}^2i_{k+2}-{i_{k}}^2i_{k+1}\big. \nonumber \\ 
\big.&-i_{k}{i_{k+1}}^2-i_{k-1}i_{k}i_{k+2}-i_{k-1}i_{k+1}i_{k+2} \big)/\Delta^{2},
\label{eq:determinant}
\end{align}
is different from zero.
Thus, the previous polynomial provides a method to analyze the time-dependent observability of the proposed model and shows that, with a proper excitation, it is possible to find an interval where the state is observable.

Regarding the possible types of excitation, it is noteworthy the case of linear evolution of $i$, i.e., $i_{k+j}=i_k+jd$
with $j \in \mathbb{N}$ and constant $d\in\mathbb{R}$. In this case, the observability can be analyzed considering that the output matrix can be expressed, for any time step, in terms of $i_k$,
\begin{equation}
C_{k+j} = \left[\begin{array}{ccc} i_k+jd & \dfrac{i_k+jd}{\Delta} & -\dfrac{i_k+\left(j-1\right)d}{\Delta} \end{array}\right],
\end{equation}
and that the $j$th power of $F$ is given by
\begin{equation}
F^j = \left[\begin{array}{ccc}1 & 0 & 0 \\[-0.5ex] 0 & j+1 & -j \\[-0.5ex] 0 & j & 1-j\end{array}\right].
\end{equation}
Then, it can be showed that, starting from the third, the $j$th row of the observability matrix is a linear combination of the two previous ones,
\begin{equation}
C_{k+j-1}F^{j-1} = 2C_{k+j-2}F^{j-2} - C_{k+j-3}F^{j-3},
\end{equation}
so that the rank of $\mathcal{O}_{\left[k,\,k+n\right]}$ is equal or less than two independently of the value of $n$.
This leads to the conclusion that no information can be extracted from intervals where ${i}$ has a linear evolution over time. {Note that steady state periods also meet this property with $d=0$.}

Let us now discuss the choice of using the inductance as state variable. Since one of the focuses of this work is to estimate the flux linkage, it might seem that \eqref{eq:coil_discrete_1} is a better choice than \eqref{eq:coil_discrete_2} to be used as output equation of the model. Actually, if the state vector is selected as 
\begin{equation}
x_k^* = \left[\begin{array}{ccc}r_k & \lambda_k & \lambda_{k-1}\end{array}\right]^\mathsf{T},
\end{equation}
an alternative output equation can be obtained from \eqref{eq:coil_discrete_1},
\begin{equation}
{y}_k = {C}_k^* {x}_k^*,
\end{equation}
\begin{equation}
C_k^* = \left[\begin{array}{ccc} i_k & 1/\Delta & -1/\Delta \end{array}\right].
\end{equation}
Then, assuming a prediction model of constant $r$ and constant increment of $\lambda$, i.e., a model whose state matrix is also given by \eqref{eq:F}, the $j$th row of the observability matrix of the alternative filter would be equal to
\begin{equation}
C_{k+j-1}^*F^{j-1} = \left[\begin{array}{ccc} i_{k+j-1} & 1/\Delta & -1/\Delta \end{array}\right].
\end{equation}
Since this shows that the second and third columns of this observability matrix are proportional, the alternative model would never be observable independently of the system excitation and of the length of the observation interval.
Consequently, we can conclude that this version of the filter is not feasible and, therefore, the selection of the inductance as state variable instead of the flux linkage is clearly justified.

{The convergence of the filter has also been studied. Considering that {\eqref{eq:state_eq}}, {\eqref{eq:output_struct}} is a time-varying discrete-time linear model, a sufficient condition for exponential stability of the filter is that the pairs $\left(F, \ G\right)$ and $\left(F, \ C_k\right)$ are, respectively, uniformly controllable and uniformly observable} \cite{anderson1981detectability}{. Since $\left(F, \ G\right)$ is time-invariant, controllability and uniform controllability are equivalent and guaranteed by the full rank of the controllability matrix $\left[G \ \ FG \ \ F^2G\right]$.
On the other hand, the pair $\left(F, \ C_k\right)$ is \textit{uniformly observable}
on the interval $t\in\left[\,k\Delta,\ (k+n)\Delta\,\right]$ if the observability Gramian,}
\begin{equation}
{W_{O}}_{\left[k,\,k+n\right]} = \sum_{i=k}^{k+n} \left(C_i F^{i-k}\right)^\mathsf{T} C_i F^{i-k},
\label{eq:gram}
\end{equation}
{satisfies, for some constants $\beta_1$ and $\beta_2$,}
\begin{equation}
0 < \beta_1 I \leq {W_{O}}_{\left[k,\,k+n\right]} \leq \beta_2 I,
\label{eq:gram2}
\end{equation}
{where $I$ is the identity matrix with proper dimensions.}
{Given {\eqref{eq:F}}, {\eqref{eq:C_k}}, and {\eqref{eq:gram}}, it is easy to see that $\beta_2$ exists whenever the current $i$ is bounded -- a condition which is always met in practice.
Then, in order to ensure the uniform observability of the system and, hence, the exponential stability of the filter, it is only necessary to check that the current excitation is such that it guarantees the existence of $\beta_1$.
Finally, it must be recalled that, since ${W_{O}}_{\left[k,\,k+n\right]} = \left(\mathcal{O}_{\left[k,\,k+n\right]}\right)^\mathsf{T} \mathcal{O}_{\left[k,\,k+n\right]}$, then}
\begin{equation}
\mathrm{rank}\left(\mathcal{O}_{\left[k,\,k+n\right]}\right) = 3 \ \ \Leftrightarrow \ \ {W_{O}}_{\left[k,\,k+n\right]} > 0,
\end{equation}
{which, together with }\eqref{eq:gram2}{, shows that observability is a necessary condition for uniform observability.}

\subsection{Algorithm equations}

The operations performed by the SEMERA estimator are summarized in Algorithm \ref{alg:algorithm}, where $\Sigma_{k/k-1}$ and $\Sigma_{k/k}$ are, respectively, the covariance matrices of the \textit{a priori} and \textit{a posteriori} state estimates. For more insight into the equations {of lines {\ref{alg:line:kalman_ini}}--{\ref{alg:line:kalman_end}}}, see the original paper by Kalman~\cite{kalman1960new} or the excellent book of Anderson and Moore~\cite{anderson2005optimal}.
{It must be noted that, when considering the probability of $x_k$ conditioned to $z_k$, the Kalman gain is obtained as $K_k = \mathrm{cov}\left(x_k,\,z_k\right)\big(\mathrm{var}\left(z_k \right)\big)^{-1}$. Thus, for the usual case of deterministic $H_k$, it is equal to $K_k = \Sigma_{k/k-1} H_k^\mathsf{T} \left(H_k \Sigma_{k/k-1} H_k^\mathsf{T} + R_k\right)^{-1}$. However, in this particular case $H_k$ is not deterministic but stochastic, so $K_k$ takes a different value. Given {\eqref{eq:kalman_output}}, {\eqref{eq:v_k2}}, and {\eqref{eq:H_k2}}, $z_k$ may be expressed as $z_k = C_k x_k + {v_\varv}_k$, which leads to
$K_k = \Sigma_{k/k-1} C_k^\mathsf{T} \left(C_k \Sigma_{k/k-1} C_k^\mathsf{T} + \sigma^2_\varv \right)^{-1}$. Since $C_k$ is not available in practice, the SEMERA algorithm computes an estimate of $K_k$, $\hat{K}_k$, by using $H_k$ instead of $C_k$,}
\begin{equation}
\hat{K}_k = \Sigma_{k/k-1} H_k^\mathsf{T} \left(H_k \Sigma_{k/k-1} H_k^\mathsf{T} + \sigma^2_\varv \right)^{-1}.
\end{equation}

{Then}, the estimates of the coil resistance and inductance, $\hat{r}$ and $\hat{l}$, are extracted from the first and second elements of the \textit{a posteriori} state estimate of the filter,
{provided that the SNRs of $\iota_k$ and $\iota_{k-1}$, which are used to calculate $H_k$, are sufficiently large.
This condition is checked in practice by a detector based on a CI
that discards, with a certain probability, that the current measurements are noise-only.
Hence, the \textit{a posteriori} estimates at step $k$ are considered valid only if
the values of $\iota_k$ and $\iota_{k-1}$ are outside the interval $\left[-n_\sigma \sigma_i,\,n_\sigma \sigma_i\right]$, where $n_\sigma$ is set according to the selected confidence.
Otherwise, the measurements are regarded as mostly noise and the estimates are calculated as $\hat{r}_k = \hat{r}_{k-1}$ and $\hat{l}_{k} = \bar{l}_0$, i.e., the resistance estimation is kept constant and the inductance is estimated to be equal to the expected initial value. This latter estimation, which may be regarded as an expert rule, is justified by the fact that non-latching electromagnetic devices always return to the initial position when the excitation is cut off. 
Consequently, the filter initial state must correspond to the resting position of the device.}
Finally, the estimate of the flux linkage is calculated as~$\hat{\lambda} = \hat{l} \iota$.

\begin{algorithm}[ht]
\caption{SEMERA algorithm.}
\label{alg:algorithm}
\begin{algorithmic}[1]
	\Require $\bar{x}_0$, $P_0$, $F$, $G$, $Q$, $\sigma^2_\varv$, $\sigma^2_i$, $\Delta$, $n_\sigma$
	\State $\hat{x}_{1/0} := \bar{x}_0$; \Comment{Initialize \textit{a priori} state estimate}
	\State $\Sigma_{1/0} := P_0$; \Comment{Initialize \textit{a priori} state covariance}
	
	\State Register $\iota_0$ and start time counter.
	\For{$k := 1$ \textbf{to} $\infty$}

		\State Wait until $t=k\Delta$; Register $u_k$ and $\iota_k$;
		\State $z_k := u_k$;
		\State $H_k := \left[\ \iota_k \ \ {\iota_k}/{\Delta} \ \ -{\iota_{k-1}}/{\Delta} \ \right]$;

		\State $\hat{K}_k := \Sigma_{k/k-1} H_k^\mathsf{T} \left(H_k \Sigma_{k/k-1} H_k^\mathsf{T} + \sigma^2_\varv \right)^{-1}$; \label{alg:line:kalman_ini}
		\State $\hat{x}_{k/k} := \hat{x}_{k/k-1} + K_k\left(z_k-H_k \hat{x}_{k/k-1} \right)$;
		\State $\Sigma_{k/k} := \left(I - K_k H_k\right) \Sigma_{k/k-1}$;

		\State $\hat{x}_{k+1/k} := F\hat{x}_{k/k}$;
		\State $\Sigma_{k+1/k} := F\Sigma_{k/k}F^\mathsf{T} + G Q G^\mathsf{T}$; \label{alg:line:kalman_end}

		\State \algorithmicif \ $\left|\iota_k\right|>n_\sigma \sigma_i \ \ \& \ \ \left|\iota_{k-1}\right|>n_\sigma \sigma_i$ 
		\State \phantom{\textbf{if}} \algorithmicthen \ $\hat{r}_k := \hat{x}_{k/k}^{(1)}$; \ $\hat{l}_{k} := \hat{x}_{k/k}^{(2)}$;
		\State \phantom{\textbf{if}} \algorithmicelse \ $\hat{r}_k := \hat{r}_{k-1}$; \ $\hat{l}_{k} := \bar{l}_0$;

		\State $\hat{\lambda}_k := \hat{l}_{k} \iota_k$;	
	\EndFor
\end{algorithmic}
\end{algorithm}

Additional estimates can be obtained if the number of turns of the coil, $N$, is known. 
First, given that the flux linkage is equal to the product of the magnetic flux through the core, $\phi$, and the number of turns of the coil, $N$,
the flux can be estimated as $\hat{\phi} = \hat{\lambda}/N = \hat{l} \iota/N$.
On the other hand, the magnetic reluctance $\mathcal{R}$ may also be estimated as $\hat{\mathcal{R}} = N^2/\hat{l}$.

\subsection{Integral Estimator}

In addition to the SEMERA algorithm, an integral {estimator} has been also developed for comparison purposes. 
The basic idea of this algorithm consists in transforming \eqref{eq:coil_cont} into integral form,
so that the flux linkage can be expressed~as
\begin{equation}
\lambda(t) = \lambda(t_0) + \int_{t_0}^t \big[\varv(\tau)-r(\tau)i(\tau)\big] \, \mathrm{d}\tau,
\label{eq:reset1}
\end{equation}
where $t_0$ is an arbitrary reference of known flux. Expressed in discrete time, it becomes
\begin{equation}
\lambda_k = \lambda_0 + \Delta\sum_{j=1}^k \big(\varv_j-r_j i_j\big),
\end{equation}
where $\lambda_0 = \lambda(t_0)$. Based on this equation, the calculation of $\lambda$ would be immediate if perfect measurements of $\varv$, $r$, and $i$ were available. However, given that only measurements of voltage and current can be obtained, 
a constant average value of resistance, $\bar{r}$, is used during the calculations instead of the time-dependent variable.
Then, replacing $\varv$ and $i$ by their respective experimental measurements, ${u}$ and $\iota$, the flux linkage
is estimated as
\begin{equation}
\hat{\lambda}_k = \lambda_0 + \Delta\left( \sum_{j=1}^k {u}_j - \bar{r} \sum_{j=1}^k \iota_j \right).
\label{eq:reset3}
\end{equation}

Since this estimate relies on {an open-loop} integration, even the slightest error in $\bar{r}$ would lead to significant cumulative errors in $\hat{\lambda}$. Thus, it becomes necessary to establish a condition in which the two integrals of the estimator are set to zero.
{Given that electromechanical devices like relays and valves operate periodically and always return to the same state at the end of the activation-deactivation cycle,
the reset event may be established at the beginning of each energizing operation, this being understood as each time the device is supplied with voltage to start the motion. Note that, at that initial point, since there is no magnetic field generated by the coil, the flux has a known constant value $\lambda_0$ which, in addition, is equal to zero for devices without permanent magnets.
Besides,} for an operation beginning at step $n$ and lasting $m$ sampling periods, the estimator should achieve $\hat{\lambda}_n = \hat{\lambda}_{n+m}$, which, using \eqref{eq:reset3}, provides {an adaptive rule} for recalculating the resistance {at the reset events},
\begin{equation}
\bar{r} = \frac{\sum_{j=n+1}^{n+m} {u}_j}{\sum_{j=n+1}^{n+m} \iota_j}.
\label{eq:reset4}
\end{equation}
Note that, since only one resistance value is obtained for each operation, the integral {estimator} cannot account for rapid variations of $r$. 
However, this should not represent a significant problem because changes in resistance are mainly due to temperature variations with slow dynamics \cite{roemer2015optimum}.

Once the estimate of the flux linkage is obtained, an estimate of the inductance is also calculated as
\begin{equation}
\hat{l}_{k} = \hat{\lambda}_k/\iota_k.
\label{eq:reset5}
\end{equation}
In order to avoid divisions by zero and prevent from 
high estimation errors when the SNR is low, {the algorithm makes use of the same CI-detector than the SEMERA estimator.
Hence, the previous expression is used at step $k$ only when the absolute values of $\iota_k$ and $\iota_{k-1}$ are higher than $n_\sigma$ times the standard deviation of the current measurement noise}. Otherwise, the inductance is considered equal to $\bar{l}_0$.
The operations performed by the integral {estimator} are summarized in Algorithm \ref{alg:integral_obs}.

\begin{algorithm}[th]
\caption{Integral {estimator}.}
\label{alg:integral_obs}
\begin{algorithmic}[1]
	\Require $\bar{r}_0$, $\bar{l}_0$, $\lambda_0$, $\sigma^2_i$, $\Delta$, $n_\sigma$
	\State $\bar{r} := \bar{r}_0$; \Comment{Initialize parameter $\bar{r}$}
	\State $S_{u} := 0$; \hspace{1ex}$S_\iota := 0$; \Comment{Initialize integrals}
	\For{$k := 1$ \textbf{to} $\infty$}
		\State Wait until $t=k\Delta$; \hspace{1ex}Register ${u}_k$ and $\iota_k$;
		\State $S_{u} := S_{u} + {u}_k$; \hspace{1ex}$S_\iota := S_\iota + \iota_k$;
		\State $\hat{\lambda}_k := \lambda_0 + \Delta\left( S_{u} - \bar{r} S_\iota \right)$;
		\State \algorithmicif \ $\left|\iota_k\right|>n_\sigma \sigma_i \ \ \& \ \ \left|\iota_{k-1}\right|>n_\sigma \sigma_i$ 
		\State \phantom{\textbf{if}} \algorithmicthen \ $\hat{l}_{k} := \hat{\lambda}_k/\iota_k$; 
		\State \phantom{\textbf{if}} \algorithmicelse \ $\hat{l}_{k} := \bar{l}_0$;					
		\State \algorithmicif \ start of energizing operation 
		\State \phantom{\textbf{if}} \algorithmicthen \ $\bar{r} := S_{u}/S_\iota$; \hspace{1ex}$S_{u} := 0$; \hspace{1ex}$S_\iota := 0$;
		\State $\hat{r}_{k} := \bar{r}$;
	\EndFor
\end{algorithmic}
\end{algorithm}

\section{Simulation}

In order to analyze the performance of the proposed estimators, a dynamic model of an electromagnetic actuator has been developed and some simulations have been carried out.

\subsection{Model equations}

\begin{figure}[tp]
\centering
\subfloat[]{\begin{minipage}[b][40.5mm][c]{4cm}\centering \includegraphics[height=40mm]{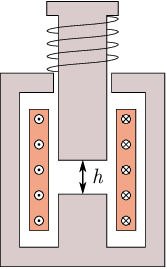} \end{minipage}\label{fig:plunger_type_actuator_a}}
\hspace{1em}
\subfloat[]{\begin{minipage}[b][40.5mm][c]{2cm}\centering \includegraphics[width=40mm,angle=270,origin=c]{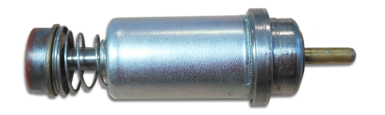} \end{minipage}\label{fig:plunger_type_actuator_b}}
\caption{(a)~Schematic diagram of a linear solenoid actuator and (b)~actual actuator (solenoid valve). The movable core is pulled towards zero gap by reluctance force. The opposite motion is driven by a spring force.}
\label{fig:plunger_type_actuator}
\end{figure}

The model presented in this section corresponds to a linear solenoid plunger-type actuator (see Fig.~\ref{fig:plunger_type_actuator}). Nevertheless, if needed, it could be easily adapted to other types of devices such as pivoted-armature or E-core actuators with minor modifications. In order to improve the readability of the equations, the explicit dependence of variables on time is omitted within the section.

As stated, the electromagnetic dynamics of the system is governed by \eqref{eq:coil_cont}. Additionally, a constitutive relation between electric current and magnetic linkage has to be established. Considering the magnetic equivalent circuit (MEC) approach, this relation is given by Hopkinson's law, $Ni=\phi\mathcal{R}$, which, considering that $\lambda = N\phi$, transforms into
\begin{equation}
N^2i = \lambda\mathcal{R}.
\label{eq:hopkinson_alt}
\end{equation}

The reluctance of the MEC, $\mathcal{R}$, can be expressed as the sum of the reluctances of the air gap, $\mathcal{R}_\mathrm{air}$, and of the iron core, $\mathcal{R}_\mathrm{iron}$. For simplicity of the model, the air reluctance is assumed proportional to the gap length, i.e., flux fringing effects are neglected.
However, since magnetic saturation should never be considered negligible in this type of devices, the iron reluctance does account for this phenomenon by means of the well-known Fr\"ohlich-Kennelly saturation model, as in \cite{ramirez2016new}. These assumptions lead to a reluctance of the form
\begin{equation}
\mathcal{R} = \mathcal{R}_\mathrm{air} + \mathcal{R}_\mathrm{iron} = k_\mathrm{air} h + \frac{{\mathcal{R}_\mathrm{iron}}_0}{1-\left|\lambda\right|/\lambda_\mathrm{sat}},
\label{eq:reluctance}
\end{equation}
where $k_\mathrm{air}$ is the proportionality constant of the air reluctance, $h$ is the gap length (see Fig.~\ref{fig:plunger_type_actuator}), ${\mathcal{R}_\mathrm{iron}}_0$ is the iron reluctance for zero flux, and $\lambda_\mathrm{sat}$ is the flux linkage saturation level. Combining \eqref{eq:coil_cont} with \eqref{eq:hopkinson_alt} and \eqref{eq:reluctance}, the dynamic equation of the flux linkage is finally obtained as
\begin{equation}
\frac{\mathrm{d}\lambda}{\mathrm{d}t} = f_\lambda\left(\lambda,h,u\right) = u - \frac{r\lambda}{N^2}\left( k_\mathrm{air} h + \frac{{\mathcal{R}_\mathrm{iron}}_0}{1-\left|\lambda\right|/\lambda_\mathrm{sat}}\right).
\end{equation}

On the other hand, the linear movement of the plunger, of mass~$m$, is directed by Newton's second law. The net force driving the motion, $F_\mathrm{total}$, is the sum of the magnetic force, the elastic force exerted by the return spring, and a damping term to account for friction forces, i.e.,
\begin{equation}
F_\mathrm{total} = F_\mathrm{mag}-k_\mathrm{s}\left(h-h_\mathrm{s}\right) -c\frac{\mathrm{d}h}{\mathrm{d}t},
\end{equation}
where $k_\mathrm{s}$ is the stiffness constant of the spring, $h_\mathrm{s}$ is the gap length for zero spring force,
$c$ is the damping coefficient and $F_\mathrm{mag}$ is given \cite{ramirez2016new} by
\begin{equation}
F_\mathrm{mag} = -\frac{1}{2}\phi^2\frac{\partial\mathcal{R}}{\partial h} = -\frac{1}{2}\frac{\lambda^2}{N^2}\frac{\partial\mathcal{R}}{\partial h} = -\frac{\lambda^2 k_\mathrm{air}}{2N^2}.
\end{equation}

\begin{figure}[tp]
\centering
\includegraphics[]{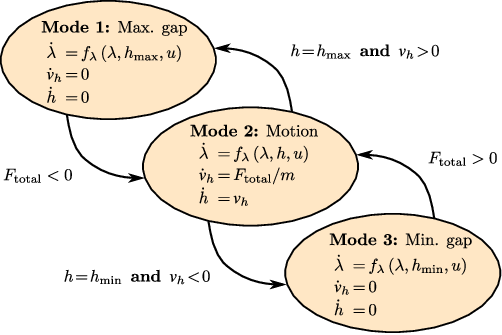}
\caption{Hybrid automaton to model the actuator dynamics. The motion of the plunger is restricted to $h\in\left[h_\mathrm{min},h_\mathrm{max}\right]$. Variable $\varv_h$ represents the velocity of the plunger along the gap direction.}
\label{fig:hybrid_automaton}
\end{figure}

Considering that the motion of the plunger is restricted by mechanical constraints, $h\in\left[h_\mathrm{min},h_\mathrm{max}\right]$, the system is modeled as a hybrid system with three dynamic modes, one for the mechanical movement and another two corresponding to the boundaries (see Fig.~\ref{fig:hybrid_automaton}).

\subsection{Simulation results and discussion}

The values of the model parameters used in the simulations are presented in Table~\ref{tab:model_parameters}. These correspond to the actual valve depicted in Fig.~\ref{fig:plunger_type_actuator_b} and have been determined by both direct inspection and experimental identification procedures.
On the other hand, the parameters used by the estimator are shown in Table~\ref{tab:filter_parameters}. 
{The variances and expected values of $r_0$ and $l_0$ have been set according to real measurements of several valves in their resting positions from an impedance analyzer. Besides, we have carried out some preliminary simulations of the actuator under a square wave input and we have analyzed the dynamic behavior of the inductance to set a proper value for $\sigma_{\ddot{l}}$. On the contrary, $r$ is not expected to have great variations in reality, so $\sigma_{\dot{r}}$ has been set to an arbitrary small value. In addition, we have analyzed some measurements from the voltage and current sensors that are used in the experiments, so the values of $\sigma_{\varv}$ and $\sigma_{i}$ used in the simulations are realistic.
The probability of the CI has been set to a conservative value of 99.9\% because we consider that, even if $\mathrm{SNR}_\iota$ is high enough, it is always preferable to use the expert rule with small values of $\iota_k$.
The sampling period, which is $\Delta=50$~$\mu$s, has been optimized by simulation to minimize the estimation error.}

\begin{table}[!tp]
\caption{Model parameters}
\label{tab:model_parameters}
\centering
	\begin{tabular}{cc}
		\hline
		& \\[-1.1em]
		{\bfseries\small Parameter} & {\bfseries\small Value}\\[0.1em]
		\hline
		& \\[-1.1em]
		$N$ & {\small $1200$}\\
		$k_\mathrm{air}$ & {\small $2.7\!\cdot\!10^{10}$ H$^{-1}$/m}\\
		${\mathcal{R}_\mathrm{iron}}_0$ & {\small $3.25\!\cdot\!10^{6}$ H$^{-1}$}\\
		$\lambda_\mathrm{sat}$ & {\small $0.024$ Wb}\\
		$m$ & {\small $1.6$ g}\\[0.1em]
		\hline
	\end{tabular}
	\hspace{1em}
	\begin{tabular}{cc}
		\hline
		& \\[-1.1em]
		{\bfseries\small Parameter} & {\bfseries\small Value}\\[0.1em]
		\hline
		& \\[-1.1em]
		
		$k_\mathrm{s}$ & {\small$37$ N/m}\\
		$h_\mathrm{s}$ & {\small$22.5$ mm}\\
		$c$ & {\small$0.4$ Ns/m}\\
		$h_\mathrm{min}$ & {\small$0$ mm}\\
		$h_\mathrm{max}$ & {\small$0.9$ mm}\\[0.1em] 
		\hline
	\end{tabular}
\vspace{\floatsep}
\caption{Filter parameters (Valve case)}
\label{tab:filter_parameters}
	\begin{tabular}{ccc}
		\hline
		& \\[-1.1em]
		{\bfseries\small Parameter} & {\bfseries\small Value}\\[0.1em]
		\hline
		& \\[-1.1em]
		$\bar{r}_0$ &{\small 77.5 $\Omega$}\\
		$\sigma_{r_0}$&{\small  1 $\Omega$}\\ 
		$\bar{l}_0$ &{\small 50 mH}\\
		$\sigma_{l_0}$ &{\small 5 mH}\\
		- &-\\[0.1em]
		\hline
	\end{tabular}\hspace{1em}\begin{tabular}{ccc}
		\hline
		& \\[-1.1em]
		{\bfseries\small Parameter} & {\bfseries\small Value}\\[0.1em]
		\hline
		& \\[-1.1em]
		$\sigma_{\dot{r}}$ &{\small 1 $\Omega$/s}\\ 
		$\sigma_{\ddot{l}}$ &{\small 10$^{8}$ H/s$^2$}\\
		$\sigma_{\varv}$ &{\small 15 mV}\\
		$\sigma_{i}$ &{\small 1 mA}\\
		$n_\sigma$ &{\small 3.29 (99.9\% CI)}\\[0.1em]
		\hline
	\end{tabular}
\end{table}

\begin{figure}[tp]
\centering
\includegraphics[]{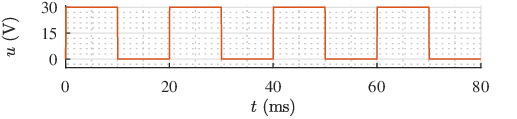}\\[0.8ex]
\includegraphics[]{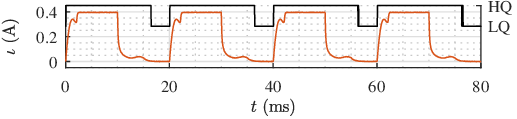}\\[0.8ex]
\includegraphics[]{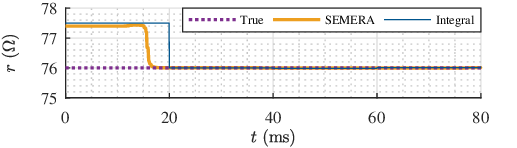}\\[0.8ex]
\includegraphics[]{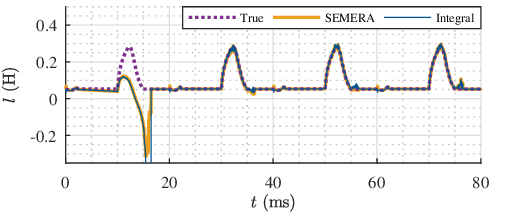}\\[0.8ex]
\includegraphics[]{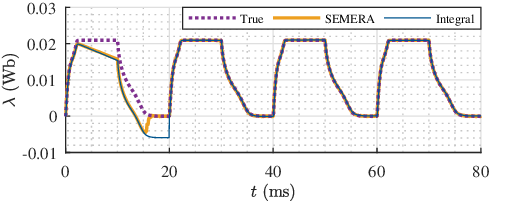}\\[0.8ex]
\includegraphics[]{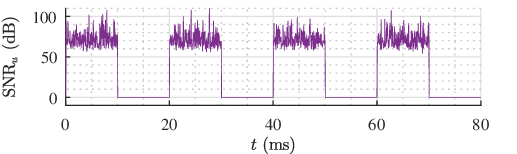}\\[0.8ex]
\includegraphics[]{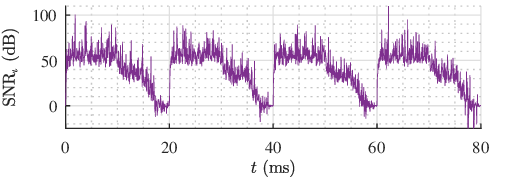}\\[0.8ex]
\includegraphics[]{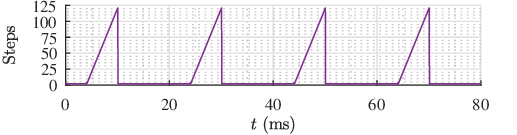}
\caption{Valve simulation results. Four activation-deactivation cycles. From top to bottom: voltage measurement, current measurement (with CI-based classification),
resistance estimation, inductance estimation, flux linkage estimation, voltage SNR, current SNR, time steps since the last observable state.}
\label{fig:simulation_1}
\end{figure}

The simulation results are presented in Fig.~\ref{fig:simulation_1} and correspond to a series of activations and deactivations of the actuator at supply voltage of 30~V. In total, four cycles of 20~ms are represented in the figures. The first two plots show respectively the simulated measurements of voltage and current, i.e., the variables used by the estimators. {The result of the CI-based noise detector, which classifies the current measurements as high-quality (HQ) or low-quality (LQ), is also represented in the second plot.} Then, the three following graphs show the estimations of resistance, inductance and flux linkage together with their respective true values. Note that the simulated value of resistance has been deliberately set to a value other than the initial value of the filters, $\bar{r}_0$, so that the transient response could be analyzed. The sixth and seventh graphs show, respectively, the SNRs of the voltage measurement and of the current measurement, which are calculated as $\mathrm{SNR}_u = 20\,\mathrm{log}_{10} \left({u}/{v_\varv}\right)$ and $\mathrm{SNR}_\iota = 20\,\mathrm{log}_{10} \left(\iota/{v_i}\right)$. Finally, the last plot represents, for each time-instant, the number of time steps since the last observable state. Note that, according to the size of the state vector, the minimum number of time steps required for a state to be observable is two.

As can be seen, the performances of the two estimators during the first activation-deactivation cycle are considerably different. Since the integral {estimator} does not modify the resistance value until the first reset event, the small errors in $\hat{r}$ (less than~2\%) lead to much higher errors (greater than 100\%) when estimating both the inductance and the flux linkage. Actually, the inductance estimation of the integral {estimator} goes far beyond the limits of the graph (it has not been completely represented for clarity reasons). On the other hand, the SEMERA algorithm has a similar behavior at the beginning, but it is able to correct the estimates during the operation and achieves much lower estimation errors, near to zero, before the end of the first cycle. This is partially due to the fact that the flux linkage is estimated through the inductance, which forces $\hat{\lambda}$ to decrease rapidly when the current measurement approaches zero.

Then, after the first cycle, the estimations present a different behavior. As can be seen, the resistance and flux linkage estimations given by both {estimators} are almost equal to the true values, so it can be concluded that the two filters achieve a very good performance with respect to these variables. {On the other hand, the inductance estimations are also very close to the true values except during two periods for each operation: a short transient after the voltage positive step ($t=\ $20, 40, and 60 ms), and a period after the current drops close to zero ($t=\ $35, 55, and 75 ms). Note that, while the first periods are intrinsic to the dynamics of the estimators, the second ones are related to a low SNR of the current measurement. In fact, it can be seen that the noisy behavior starts when $\mathrm{SNR}_\iota$ falls approximately below 30 dB, and that the problem is later detected and overcome by means of the CI-based detector, which acts approximately for $\mathrm{SNR}_\iota<20$~dB.}

It is also noteworthy that there is no need to design a specific activation signal to provide observability; the standard square-wave usually employed to activate these devices provides minimum-time observability except during the steady-state periods. {In this regard, note that the current being constant is simply a particular case of linear evolution over time, $i_{k+j}=i_k+jd$, with $d=0$. Hence, the results are} in accordance with the observability analysis presented in Section~\ref{subsec:observability}.

Finally, the root-mean-square errors (RMSE) of the estimates during the simulation have been calculated and are presented in~Tables~\ref{tab:estimation_error} and~\ref{tab:estimation_error_2}. {Regarding the errors during the first operation (Table~{\ref{tab:estimation_error}}), it is showed that the SEMERA estimator performs better, particularly for the inductance estimation. Then, after the first cycle, once the estimators have converged, the errors (Table~{\ref{tab:estimation_error_2}}) are one or two orders of magnitude smaller, but in any case the SEMERA performance is still better for the three variables.}

\begin{table}[!t]
	\caption{Estimation errors during the first operation ($t< 20 \ \mathrm{ms}$).}
	\label{tab:estimation_error}
	\centering
	\begin{tabular}{lccc}
		\hline
		& \\[-1em]
		{\bfseries\small Algorithm} & {\bfseries\small RMSE\,($\,\hat{r}\,$)} & {\bfseries\small RMSE\,($\,\hat{l}\,$)} & {\bfseries\small RMSE\,($\,\hat{\lambda}\,$)}\\[0.1em]
		\hline
		& \\[-1em]
		{\small SEMERA} & {\small $1.244$ $\Omega$} & {\small $0.1022$ H} & {\small $3.602$ mWb} \\
		{\small Integral} & {\small $1.500$ $\Omega$} & {\small $0.2512$ H} & {\small $4.645$ mWb} \\
		{\small \textit{Ratio} (S/I)} & {\small \textit{0.8296}} & {\small \textit{0.4069}} & {\small \textit{0.7756}}\\[0.1em] 
		\hline
	\end{tabular}
\vspace{\floatsep}
	\caption{Estimation errors after the first operation ($t> 20 \ \mathrm{ms}$).}
	\label{tab:estimation_error_2}
	\centering
	\begin{tabular}{lccc}
		\hline
		& \\[-1em]
		{\bfseries\small Algorithm} & {\bfseries\small RMSE\,($\,\hat{r}\,$)} & {\bfseries\small RMSE\,($\,\hat{l}\,$)} & {\bfseries\small RMSE\,($\,\hat{\lambda}\,$)}\\[0.1em]
		\hline
		& \\[-1em]
		{\small SEMERA} &{\small  $4.199$ m$\Omega$} &{\small  $5.022$ mH} &{\small  $0.1136$ mWb} \\
		{\small Integral} &{\small  $10.30$ m$\Omega$} &{\small  $5.158$ mH} &{\small  $0.1445$ mWb} \\
		{\small \textit{Ratio} (S/I)} &{\small  \textit{0.4077}} &{\small  \textit{0.9735}} &{\small  \textit{0.7866}}\\[0.1em] 
		\hline
	\end{tabular}
\end{table}

\section{Experimental Evaluation}

\begin{figure}[tp]
\centering
\subfloat[]{\begin{minipage}[b][30mm][c]{40mm}\centering \includegraphics[]{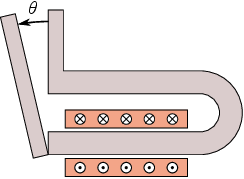} \end{minipage}\label{fig:pivoted_armature_actuator}}
\hspace{1em}
\subfloat[]{\begin{minipage}[b][30mm][c]{4cm}\centering \includegraphics[width=40mm]{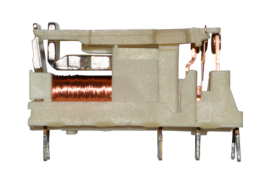} \end{minipage}}
\caption{SPDT power relay. (a) Schematic diagram of the reluctance actuator and (b) actual relay.}
\label{fig:relay}
\end{figure}

Once the performance of the estimators has been studied and compared by simulations, in this section we analyze their operation under real conditions. For this purpose, both filters have been implemented on a low-cost ARM-Cortex M3 microcontroller and tested on two different devices: the solenoid valve presented in the previous section (see Fig.~\ref{fig:plunger_type_actuator_b}) and a single-pole double-throw (SPDT) power relay based on a pivoted-armature actuator (see Fig.~\ref{fig:relay}). Both devices have been activated and deactivated periodically at supply voltage of 30~V, as in the simulations and as they are usually operated. When applied to the valve, the parameters employed by the filters are those already presented in Table~\ref{tab:filter_parameters}. On the other hand, Table~\ref{tab:filter_parameters_relay} shows the parameters for the case of the power relay, which presents very different values of inductance and resistance. The value of $\sigma_i$ is also different because a different current sensing method has been used on this latter device. {The sampling period, which is $\Delta\!=\!50$ $\mu$s as in the simulations, is enough to run both algorithms in 32 bit floating point (time per iteration, approximately, SEMERA: 38 $\mu$s, Integral: 2 $\mu$s).}

In addition, since the true values of resistance, inductance and flux linkage are not accessible in reality, an offline non-causal version of the integral {estimator} has been also implemented to provide a deeper analysis. Unlike the online version, which uses data of each operation to recalculate the resistance and estimate the variables of the following one, this estimator firstly computes the resistance of each and every operation and then estimates the rest of the variables. Hence, although it also assumes a constant value between reset events, the most accurate value possible of $\bar{r}$ is utilized.

\begin{table}[t]
\caption{Filter parameters (Relay case)}
\label{tab:filter_parameters_relay}
\centering
	\begin{tabular}{ccc}
		\hline
		& \\[-1.1em]
		{\bfseries\small Parameter} & {\bfseries\small Value}\\[0.1em]
		\hline
		& \\[-1.1em]
		$\bar{r}_0$ &{\small  1560 $\Omega$}\\
		$\sigma_{r_0}$ &{\small  100 $\Omega$}\\ 
		$\bar{l}_0$ &{\small  1 H}\\
		$\sigma_{l_0}$ &{\small  250 mH}\\
		- &-\\[0.1em]
		\hline
	\end{tabular}\hspace{1em}\begin{tabular}{ccc}
		\hline
		& \\[-1.1em]
		{\bfseries\small Parameter} & {\bfseries\small Value}\\[0.1em]
		\hline
		& \\[-1.1em]
		$\sigma_{\dot{r}}$ &{\small  20 $\Omega$/s}\\ 
		$\sigma_{\ddot{l}}$ &{\small  $5\!\cdot\!10^{9}$ H/s$^2$}\\
		$\sigma_{\varv}$ &{\small  15 mV}\\
		$\sigma_{i}$ &{\small  0.05 mA}\\
		$n_\sigma$ &{\small 3.29 (99.9\% CI)}\\[0.1em]
		\hline
	\end{tabular}
\end{table}

\begin{figure}[!t]
\centering
\includegraphics[]{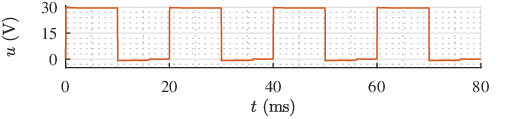}\\[1ex]
\includegraphics[]{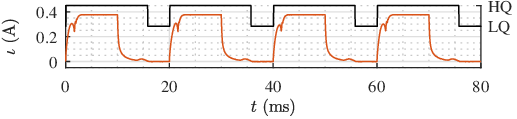}\\[1ex]
\includegraphics[]{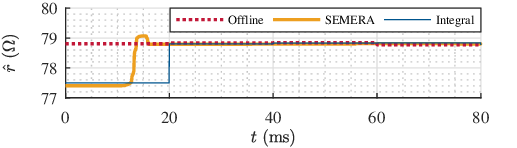}\\[1ex]
\includegraphics[]{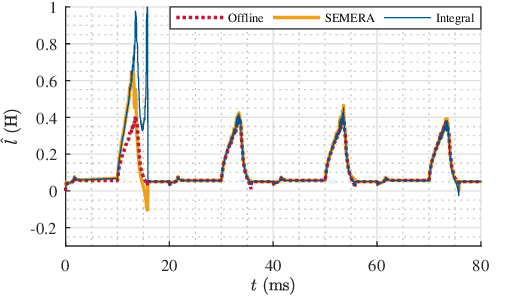}\\[1ex]
\includegraphics[]{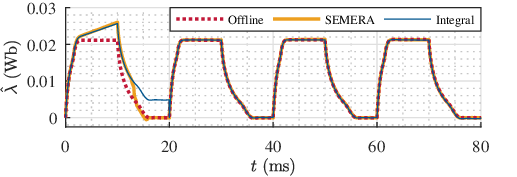}
\caption{Valve experimental results. From top to bottom: voltage measurement, current measurement (with CI-based classification), resistance estimation, inductance estimation, flux linkage estimation.}
\label{fig:test_1}
\end{figure}

\begin{figure}[!t]
\centering
\includegraphics[]{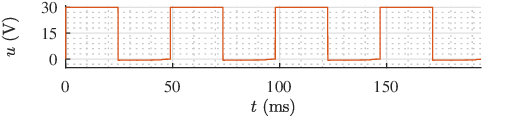}\\[1ex]
\includegraphics[]{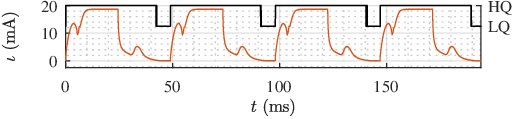}\\[1ex]
\includegraphics[]{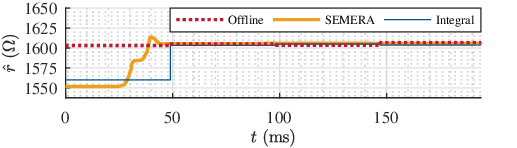}\\[1ex]
\includegraphics[]{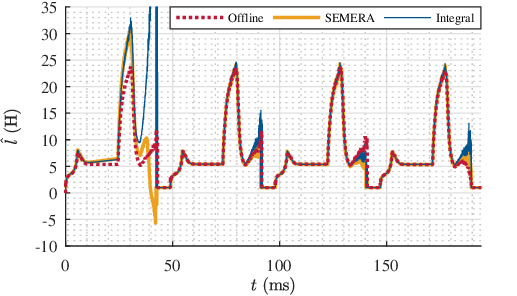}\\[1ex]
\includegraphics[]{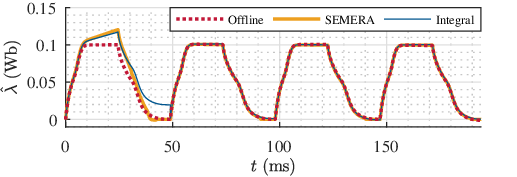}
\caption{Relay experimental results. From top to bottom: voltage measurement, current measurement (with CI-based classification), resistance estimation, inductance estimation, flux linkage estimation.}
\label{fig:test_3}
\end{figure}

The results corresponding to the valve and the relay are respectively presented in Figs.~\ref{fig:test_1} and~\ref{fig:test_3}. Considering the offline estimates as the most accurate, it can be seen that the dynamics of the online estimations are very close to the simulation results already presented. It is showed that the highest errors occur during the first activation-deactivation cycle, when the small resistance estimation error leads to high errors in $\hat{l}$, although the SEMERA estimations converge much faster to the true values. The two estimators behave similarly once the first cycle has finished; they provide very good estimations of $r$, $l$, and $\lambda$. {The noisy behavior of $\hat{l}$ during the periods of low $\mathrm{SNR}_\iota$, which has been already observed in the simulations, can also be noticed here in the relay test (around $t=$ 85, 135, and 185 ms), although it is almost unnoticeable in the valve experiment. Nevertheless, the CI-based classifier is able to detect the problem and the expert rule corrects the estimation when $\mathrm{SNR}_\iota$ is very low.}

Regarding the evolution of the variables, it can be firstly seen that the resistances of both devices keep an almost constant value during the experiments, which is the expected behavior. The same applies to the flux linkage, which oscillates between zero and a maximum steady value. Finally, the inductance behavior shows that, as already stated in the model section, this variable depends both on the position of the mechanism and on the magnetic flux. In this regard note that, if the inductance only depended on the position, it would oscillate strictly between two values corresponding to the bounds of the motion, which is clearly not the case.


\begin{figure}[tp]
\centering
\includegraphics[]{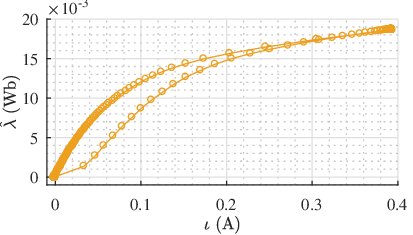}
\caption{Path-dependent relation between flux linkage and electric current. The behavior is due to both magnetic hysteresis and eddy~currents.}
\label{fig:hyst}
\end{figure}

In order to provide more insight into the relation between electric current and magnetic flux, an additional experiment has been performed with the valve. The plunger has been locked at the zero-gap position and the coil has been supplied with a 30~V square wave as in the previous tests. Then, the flux linkage has been obtained by means of the SEMERA algorithm and the results, once the estimation has converged, have been represented in the $\lambda$--$i$ plane (see Fig.~\ref{fig:hyst}). Two conclusions can be drawn from the graph: one, that magnetic saturation exists and has a great impact in the dynamics of the device, and second, that the relation between $i$ and $\lambda$ for any given position is not static but depends on past values. This, as explained in some previous works~\cite{vrijsen2014prediction,braun2015semilinear}, is due both to magnetic hysteresis and to eddy currents.

\section{Conclusions}

In this paper, we have presented a novel stochastic observer for reluctance actuators, SEMERA, which is able to estimate the magnetic linkage and the time-variant electrical parameters of these devices, i.e., the resistance and the inductance, even under temperature variations and measurement noise. As has been shown, the algorithm has proved to be highly accurate and able to handle long unobservable periods and poor SNRs. {Besides, the experimental results show that it is fully applicable to any reluctance actuator, independently of the shape, the materials, or the mechanical design, because it only relies on the electrical equation of a variable inductor. In this regard, it is much more versatile than other model-based estimators recently presented.}

Additionally, we have also designed an ad hoc integral {estimator} to provide a comparative analysis. This estimator, which makes use of the repetitive operating mode of valves and switches, has also showed a good precision while requiring simpler calculations. However, an application problem may be encountered with this latter approach: if the time between operations is long enough that the temperature changes considerably, the corresponding change in the electrical resistance may lead to high transient errors that will not be reduced until the end of a complete operation.

Apart from the aforementioned variables, additional estimates might be derived from the SEMERA observer. The resistance may be used, e.g., to estimate the temperature of the device or to detect faults, and the magnetic linkage estimation allows for characterizing the dynamic behavior between flux and current. In addition, it is known in the literature that the inductance is related to the position of the device and, consequently, that it may be used to control its motion. Nevertheless, we have showed that, prior to performing the estimation, it is imperative that the phenomena of magnetic hysteresis and eddy currents are included into the models.

\bibliographystyle{IEEEtranTIE}

\end{document}